\documentclass[journal]{IEEEtran}

\usepackage{cite}
\usepackage{amsmath,amssymb}
\usepackage{algorithm}
\usepackage{algorithmic}
\usepackage{graphicx}
\usepackage{tikz}
\usetikzlibrary{arrows.meta,positioning}
\usepackage{booktabs}
\usepackage{multirow}
\usepackage{xcolor}
\usepackage{url}
\usepackage[hidelinks]{hyperref}
\usepackage{xspace}

\newcommand{\sys}{\textsc{PlaceMark}\xspace}

\begin{document}

\title{Blind Watermarking of Software Repositories\\via Module Placement}

\author{Khadija~Mohamed \quad Khaled~Fathy \quad Mamdouh~Farouk \quad Mohamed~M.~Darwish\\[3pt]
\normalfont\normalsize Department of Computer Science, Faculty of Computers and Information\\
\normalfont\normalsize Assiut University, Assiut, Egypt\\[2pt]
\normalfont\normalsize khadija.mohamed@aun.edu.eg \quad khussain@aun.edu.eg\\
\normalfont\normalsize mamdouh@fci.au.edu.eg \quad mohamed\_darwish@aun.edu.eg}

\markboth{}{}

\maketitle

\begin{abstract}
Source-code watermarking supports provenance tracking by embedding a recoverable identifier in a
software artifact. Many existing approaches encode this identifier in tokens or program syntax,
exposing the watermark to transformations that modify the source text. We present \sys{}, a blind
watermarking system that encodes a payload in the placement of modules within a repository. Each
eligible module is identified by its filename and top-level definition names. A secret key
determines its ordered candidate destinations and its position in the reading order; the module's
location within its candidate list determines the encoded symbol. Embedding relocates selected
modules and updates their import references. Extraction uses the distributed repository, the key,
and agreed decoding parameters, without an original copy or placement ledger. Capacity measurements
on ten Python repositories yield nominal capacities of $249$ to $2{,}166$ bits in nine
repositories, while one provides no placement channel. In $216$ trials of the evaluated text
transformations, the payload is recovered exactly. Relocation degrades recovery gradually on \textsc{Django}, \textsc{Airflow} and \textsc{Salt},
while on \textsc{Pylint} a path-based exclusion in our implementation lets relocation change carrier
membership, and we
measure it for an untargeted adversary and for one that concentrates its budget on the public
carrier set. Changes to filenames or definition names alter the identity input and are measured
separately. Exact recovery also depends on placement feasibility. The results characterize module
placement as a carrier that can complement watermarks embedded in source text.
\end{abstract}

\IEEEpeerreviewmaketitle


\section{Introduction}

Source code can be copied, modified, and redistributed with little effort, making its provenance
difficult to establish once it leaves the original repository. A project may be republished under
another name or incorporated into a different product after changes that obscure its origin. This
motivates a practical question: can a distributed repository retain a recoverable identifier
without relying on the cooperation of the party redistributing it?

Watermarking addresses this need by embedding a recoverable mark within the artifact itself. For
source code, the carriers reported in the literature lie within the code representation: a recent
systematic review classifies existing schemes entirely at the level of tokens, syntax, and the
abstract syntax tree~\cite{fahad2026review}. Identifiers are renamed according to a secret
rule~\cite{srcmarker}, statements are replaced with semantically equivalent
variants~\cite{clasp,mcss,swarl}, or, for model-generated code, the token distribution is biased
during generation~\cite{sweet,kirchenbauer2023}. These schemes can be effective under the transformations considered in their evaluations, but they
operate on source representations that maintenance and redistribution routinely modify, so their
robustness depends on which elements an attack changes. The Transformer variant of
\textsc{SrcMarker}, for example, reports $97.26\%$ bit accuracy without attack and $62.68\%$ after
full variable substitution~\cite{srcmarker}. This motivates a complementary carrier without
implying that all source-text watermarks respond identically to the same transformation.

We observe that a repository offers a second degree of freedom, one that, to the best of our
knowledge, existing source-code watermarking schemes do not use: the arrangement of its modules and
files. Module placement in a software project is often flexible. A module implementing a caching
helper, for example, may reasonably be placed in a utilities package, beside its principal consumer,
or in a core package, with the choice reflecting maintainer judgement rather than a functional
constraint. That choice can therefore carry information. If a module can be placed in any of eight
packages without affecting program behaviour, its placement encodes three bits, and a repository
containing hundreds of such modules provides a channel of hundreds to thousands of bits. Writing to
this channel inserts no code and deletes none. Its edits are confined to the references that name a
module it moves, which are the edits a maintainer would make by hand for the same move. The message is encoded in the arrangement of the files themselves.

Blind extraction creates a synchronisation problem. Without the original repository or a placement
ledger, the decoder must identify the same carriers and read them in the same order as the
encoder. A path-dependent eligibility rule may violate this condition after relocation: a moved
file can enter or leave the carrier set, shifting every subsequent symbol. The decoder must also
reconstruct the same destination universe, since a change there can alter the candidate lists and
the radix. We formalise this requirement, show how it constrains the identity function and the
reading order, and measure the consequences of violating it.

We instantiate these principles in \sys{}, a watermarking system for repository structure. Each
carrier file is identified by its own content, its filename together with the names it defines, so
that it remains recognisable wherever it is placed. A keyed pseudorandom function assigns each
carrier an ordered list of packages it may legally occupy, and the index of its actual package
within that list encodes the symbol it carries. A second keyed permutation determines the order in
which carriers are read, preventing the payload-bearing subset from being identified without the
key. Files whose relocation could alter program behaviour, such as those discovered by frameworks
through directory enumeration, are excluded using a test based on the file's own syntax tree, which
remains unchanged by relocation.

We census ten production repositories and run the experiments on a subset of them. The channel
provides a nominal capacity of up to $2{,}166$
bits, and recovers the payload exactly when the selected positions are writable. Across $216$
measured trials covering renaming of up to $100\%$ of local variable names, dead-code insertion,
exception wrapping, print insertion, and their combination, no payload bit was lost. Under relocation, the
transformation to which a placement watermark is inherently exposed, degradation is gradual rather
than abrupt. The placement carrier and the content carriers used by prior work are therefore
complementary.

\medskip\noindent\textbf{Contributions.}
\begin{enumerate}
\item \textbf{Watermarking by module placement, extracted blindly.} We introduce a watermarking
scheme that encodes the payload in the placement of source files within the repository hierarchy. To
the best of our knowledge, this is the first source-code watermark that uses module placement as its
carrier. The payload is represented by the arrangement of the repository, without inserting or
deleting code lines to carry the watermark, and is recovered using only the distributed repository
and a secret key, with no reference copy, trained model, or stored side information.

\item \textbf{Relocation-invariant carrier selection.} We formalise the requirement that the
carrier set and reading order recomputed by a blind decoder from the artifact be invariant under the
transformation the watermark is designed to survive. We show that this requirement constrains the identity function and
the carrier reading order, and verify empirically on \textsc{Django} and \textsc{Salt} that the
resulting carrier set remains unchanged
when half of a repository is relocated.

\item \textbf{Robustness to content-level removal transformations.} We evaluate \sys{} against the
code transformations used to assess the removal of existing content-carrier watermarks. Across the
$216$ measured trials on three repositories, covering renaming of up to $100\%$ of local variable
names, dead-code insertion, exception wrapping, print insertion, and their combination, no payload
bit was lost. These transformations leave the file names and top-level definition names that carrier
identity is built from unchanged.

\item \textbf{A channel of hundreds to thousands of bits.} We measure the nominal capacity on ten
production repositories, obtaining between $249$ and $2{,}166$ bits on the nine that have any
capacity at all. The results show that capacity is determined primarily by the architectural
structure that provides valid carrier placements, rather than directly by repository size.
\end{enumerate}

\section{Related Work}

\textbf{Ordering as a carrier.} The idea of encoding a payload in the arrangement of program
elements dates to the first static software watermark, which permutes basic
blocks~\cite{davidson1996}; later work permutes constant-pool entries, expression operands, and
instruction schedules~\cite{hamilton2011,anckaert2004}. \textsc{SoftMark}~\cite{softmark} is the
closest modern relative: it encodes an identifier in the order of functions within a compiled binary,
adds no code, and reports $382$ to $21{,}326$ bits per binary. \sys{} differs in substrate and in what
that substrate buys. A binary's function order is destroyed by recompilation and is not a decision any
developer makes; a module's package is a design choice that is preserved by any distribution that
keeps the source layout, including a clone, and is defensible to a reviewer. \textsc{SoftMark} also extracts with the help of a stored ledger of function locations, where
\sys{} extracts from the key and the agreed decoding parameters, with no record of where anything
was put.

The robustness literature for this family is largely negative: reordering watermarks are reported as
destroyed by any renormalisation of the order, and an evaluation of fourteen static schemes under
semantics-preserving transformation found the family broadly vulnerable to the transformations it
applied~\cite{hamilton2010eval}. Quantitative evaluation is not absent,
\textsc{SoftMark} for instance reporting a probability curve for attack success~\cite{softmark};
what we have not found is a graded recovery curve against a budget of relocations. The reason the placement carrier can offer one is that its admissible set is defined by what a
maintainer would accept, so an adversary cannot renormalise it the way a basic-block order can be
sorted; whether some canonical arrangement could be imposed while keeping the project working is an
open question rather than something we rule out.

A related idea appears in operand reordering. \textsc{Shirali-Shahreza} is reported to make such a
watermark \emph{blind}, that is extractable without an original copy or a stored record of what the
encoder did, by defining a canonical order on the operands of a swappable operation derived from
their own content~\cite{shirali2008}. That is the device \sys{} applies at file granularity: a
decoder that recomputes which files carry the payload, and the order in which they are read, from
what each file contains needs no record of where the encoder put it. The same problem has a long history in
image watermarking, where feature points computed from image content let a blind decoder resynchronise
after a geometric transform has moved everything~\cite{bas2002}. Relocation is the discrete analogue, and the property that stands in for geometric invariance is
that the decoder's reconstruction be unchanged by it, which \S\ref{sec:carriers} states
formally.

\textbf{Constraint-based watermarking.} Encoding a signature by choosing among the legal solutions of
a constrained problem was introduced for hardware intellectual property~\cite{kahng1998,qu1998} and
applied to register allocation~\cite{myles2003}. \sys{} is an instance of that framework in a new
domain: a module with several legal packages is a variable with several legal values. Placement itself
has been used as the degree of freedom in that setting, in a recent scheme that embeds a signature in
the positions assigned to cells during the physical design of an integrated circuit~\cite{pdmarks}.
The analogy is informative but the substrate is not transferable: cell coordinates are produced by a
placement tool from a netlist, not chosen by an author, and the artifact that reaches the adversary is
a layout rather than a repository that must still build and import.

\textbf{Program-internal structure as a carrier.} A second line of work encodes the payload in the
structure of the program rather than in its text: the topology of the control-flow
graph~\cite{venkatesan2001}, or a graph built in the heap and traversed at run
time~\cite{collberg1999}. These are structural watermarks, but the structure they use is internal to
the program and is reached by inserting or altering code, so the mark is again carried by the code
text and is exposed to the transformations that rewrite it. \sys{} uses structure that lies outside
the program: the arrangement of its files, which carries no statements of its own and which the
text-level transformations evaluated here do not alter.

\textbf{Source code watermarking.} Contemporary schemes carry the payload in the code text.
\textsc{SrcMarker}~\cite{srcmarker} embeds four bits per function through learned selection of
semantics-preserving transformations, and reports $97.3\%$ bit accuracy undisturbed, falling to
$62.7\%$ when every variable is renamed. Dataset-level schemes~\cite{codemark,coprotector} poison a
training corpus so that a model trained on it reproduces a trigger, and are verified statistically
against the trained model rather than extracted from an artifact; a recent purification attack removes
them with full recall in the scenarios it evaluates~\cite{decoma}. Watermarks for model-generated code bias token selection at
generation time~\cite{sweet,kirchenbauer2023}, and lose most of their detection power under dead-code
insertion or wrapping~\cite{llmcoderobust}. The operations that damage these carriers leave placement
untouched, and the operation that damages placement is not part of their attack model.

\textbf{Steganography.} When a carrier cannot be moved, the encoder knows which positions it had to
skip and the decoder does not. That asymmetry is the non-shared selection channel wet paper codes
address~\cite{fridrich2005wet,fridrich2005efficient}, and the appropriate
measures of a scheme like ours, such as embedding efficiency and the payload that can be carried as
the cover grows~\cite{filler2011stc,ker2008srl}, come from that literature. Structural carriers below
the file abstraction, such as file-system allocation order or directory-entry
order~\cite{anderson1998stegfs,mcdonald1999stegfs,aycock2015fat}, hide data in a disk rather than in a
project: they do not survive re-serialisation by an archiver or a version control system, and they do
not travel with the artifact. A placement watermark lives in tracked structure and is reconstructed
identically by every clone.

\section{Problem Statement and Threat Model}

\subsection{Setting}

A repository is a set of modules organised into packages. Writing $M$ for the modules and $P$ for
the packages, a \emph{placement} is a map $\pi : M \to P$, and the repository as shipped realises one
particular placement. Not every placement is legal: moving a module may create an import cycle,
collide with a file of the same name, or change program behaviour. Let $L(m) \subseteq P$ denote the
packages that module $m$ may occupy without any of these effects. The admissible configurations number at most $\prod_{m} |L(m)|$, with equality only if the
placements are independent, which they are not in general: a cycle or a name collision couples one
module's choice to another's. The channel therefore holds at most $\sum_{m} \log_2 |L(m)|$ bits,
an upper bound rather than a count. This is analogous to the quantity that basic-block reordering reports
as $\log_2 n!$~\cite{davidson1996,hamilton2011} and that constraint-based hardware watermarking
reports as bits per device~\cite{kahng1998}, which makes capacities across these substrates
comparable.

An embedding is a change of placement, $\pi \rightarrow \pi'$, chosen so that a keyed decoding of
$\pi'$ yields the intended message. No file is created or deleted, and no line is edited in order to
carry the payload; the only content edits an embedding performs are the reference rewrites
that keep the repository consistent after a module moves.

\subsection{Design goals}
\label{sec:requirements}

We use the standard vocabulary of the watermarking literature.
\begin{itemize}
\item \emph{Effectiveness}: the message is recovered from an unmodified marked repository. Recovery
is exact when every selected payload position is writable; a position the encoder cannot write stays
unwritten, and because the decoder does not know which positions those are it reads them as ordinary
symbols, which can leave residual errors as utilisation rises (\S\ref{sec:eval:effectiveness}).
\item \emph{Blindness}: extraction takes the marked repository, the key and the agreed decoding
parameters, and nothing else, with
no original copy, trained model, or stored side information.
\item \emph{Capacity}: how many bits a repository holds.
\item \emph{Robustness}: the message survives modification of the artifact.
\item \emph{Behaviour preservation}: the marked software does what the unmarked software did. For a
placement carrier this is the requirement most at risk, because a module that is moved out of a
directory a framework enumerates acquires or loses meaning, so the design devotes a selection rule
to it (\S\ref{sec:carriers}).
\end{itemize}

\subsection{Threat model}

We assume Kerckhoffs's principle: the adversary knows the scheme in full, including the eligibility
rule, the destination analysis and the code, and lacks only the key.

The adversary of interest is a \emph{remover}, which obtains a marked repository and wants to
redistribute it without the mark, while keeping it usable. Its budget is the fraction of the
repository it is willing to rearrange, since rearrangement is not free: it must be reviewed, it
conflicts with concurrent work, and past some level it is indistinguishable from rewriting the
project. We evaluate budgets from $5\%$ to $50\%$ of all modules.

Two attacks are outside the model. An adversary
holding the same key can overwrite the payload with its own; this scheme does not establish
ownership (\S\ref{sec:discussion}). A full rewrite of the repository lies outside this model and would
remove the structural carrier considered here.

\section{\sys{} Design}
\label{sec:design}

\begin{table}[t]
\caption{Notation used throughout Sections \ref{sec:design} and \ref{sec:eval}.}
\label{tab:notation}
\centering
\small
\begin{tabular}{ll}
\toprule
Symbol & Meaning \\
\midrule
$C$       & carrier set: modules eligible to carry payload \\
$U$       & destination universe: packages a carrier may enter \\
$\varrho$ & radix: largest power of two with $\varrho \le |U|$ \\
$E$       & symbol width in bits, $E \le \log_2 \varrho$ \\
$r$       & repetition factor: copies stored of each payload bit \\
$k$       & the secret key \\
$\mathrm{mid}(m)$ & content-derived identity of module $m$ \\
$a(c)$    & keyed candidate shortlist of $c$, $|a(c)| = \varrho$ \\
\bottomrule
\end{tabular}
\end{table}

\subsection{Overview}

\begin{figure}[t]
\centering
\begin{tikzpicture}[
  font=\scriptsize,
  box/.style={draw, rounded corners=1pt, minimum height=6.2mm, align=center, inner sep=2pt},
  lbl/.style={font=\scriptsize\itshape, text=black!65},
  ar/.style={-{Latex[length=1.4mm]}, draw=black!70}]

\node[box, minimum width=17mm] (repo) at (0,0) {repository};
\node[box, minimum width=9mm] (key) at (0,-0.95) {key $k$};

\node[box, minimum width=15mm] (elig) at (2.35,0) {eligibility\\{\tiny(content-derived)}};
\node[box, minimum width=15mm] (id)   at (4.55,0) {identity\\{\tiny leaf $\|$ defs}};
\node[box, minimum width=15mm] (cand) at (6.75,0) {candidates\\{\tiny keyed, radix $\varrho$}};

\draw[ar] (repo) -- (elig);
\draw[ar] (elig) -- (id);
\draw[ar] (id) -- (cand);
\draw[ar] (key.east) -- ++(0.35,0) |- (cand.south);

\node[box, minimum width=15mm] (sym)  at (6.75,-1.5) {symbol\\{\tiny $i \bmod 2^{E}$}};
\node[box, minimum width=15mm] (ord)  at (4.55,-1.5) {reading order\\{\tiny keyed}};
\node[box, minimum width=15mm] (msg)  at (2.35,-1.5) {payload};

\draw[ar] (cand) -- (sym);
\draw[ar] (sym) -- (ord);
\draw[ar] (ord) -- (msg);
\draw[ar] (key.south) |- ++(0.2,-0.55) -| (ord.south);

\node[lbl] at (4.00,-2.3) {identity from content; destinations from package structure};
\draw[draw=black!35, dashed, rounded corners=2pt] (1.50,-0.38) rectangle (5.40,0.42);
\draw[draw=black!35, dashed, rounded corners=2pt] (3.70,-1.88) rectangle (5.40,-1.08);
\end{tikzpicture}
\caption{\sys{} embedding and extraction. The encoder moves modules until the symbol sequence
spells the payload; the decoder recomputes the same quantities from the tree in front of it. The
dashed regions mark what Eq.~\eqref{eq:invariance} constrains.}
\label{fig:overview}
\end{figure}
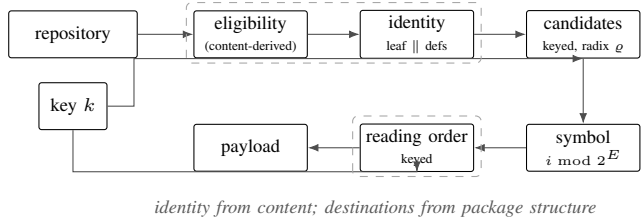

Embedding and extraction run the same four steps, illustrated in Fig.~\ref{fig:overview}. Given a
repository and a key, \sys{} first determines which modules are eligible to carry payload. It gives each eligible module an identity that does not depend on where it sits. It derives
from the key an ordered shortlist of candidate destinations per module, in which the index of a
module's own package is the symbol that module carries. Finally, it derives from the key the order in which
modules are read.

The encoder leaves a carrier in place when its current package already encodes the required symbol,
and moves it otherwise; it does not search for a globally minimal set of moves. Where every selected
position was writable, the result is a tree whose decoded symbol sequence equals
the message. The decoder repeats the same four steps on the tree in front of it and reads the
symbols off.

Algorithms~\ref{alg:embed} and~\ref{alg:extract} state both sides. The two reconstruct the carrier
set, the destination universe, the radix and the reading order by the same procedures, and that
identity is the core of the correctness argument. The decoder
reconstructs the carrier set, the radix, and the reading order from the repository and the key,
without being told any of them. The scheme's only shared state is the key $k$, the symbol
width $E$, the repetition factor $r$, and the payload length.

\begin{algorithm}[t]
\caption{\sys{} embedding}
\label{alg:embed}
\begin{algorithmic}[1]
\STATE \textbf{input:} repository $R$, key $k$, payload $w$ of $\ell = |\textsc{Bits}(w)| > 0$ bits,
       width $E \ge 1$, repetition $r$ odd and positive
\STATE $C \leftarrow \textsc{Eligible}(R)$ \COMMENT{\S\ref{sec:carriers}; one path-based exception}
\STATE $U \leftarrow \textsc{AdmissibleDestinations}(R)$
\STATE \textbf{abort} if $|U| < 2$ \COMMENT{no destination to vary}
\STATE $n \leftarrow \lfloor \log_2 |U| \rfloor$; \ $\varrho \leftarrow 2^{n}$ \COMMENT{radix}
\STATE \textbf{abort} if $E > n$ \COMMENT{width exceeds the radix}
\STATE $O \leftarrow \textsc{Sort}\big(C, \ \text{key}{=}\mathrm{PRF}_k(\text{``order''}, \mathrm{mid}(\cdot))\big)$
\STATE $b \leftarrow \textsc{Bits}(w)$ concatenated $r$ times, right-padded with zeros to a
       multiple of $E$; \ $s \leftarrow \textsc{Pack}(b, E)$
\STATE \textbf{abort} if $|s| > |O|$ \COMMENT{capacity exceeded}
\FOR{$j \leftarrow 0$ \textbf{to} $|s|-1$}
  \STATE $c \leftarrow O[j]$
  \STATE $a \leftarrow \textsc{Sort}\big(U, \text{key}{=}\mathrm{PRF}_k(\text{``cand''}, \mathrm{mid}(c), \cdot)\big)[0{:}\varrho]$
  \STATE $K \leftarrow \{\, a[i] \ : \ i \bmod 2^{E} = s_j \,\}$ \COMMENT{equivalence class}
  \IF{$\mathrm{pkg}(c) \in K$}
    \STATE \textbf{continue} \COMMENT{already correct: no move}
  \ENDIF
  \STATE $F \leftarrow \{\, p \in K : \text{no import cycle, no name collision} \,\}$
  \IF{$F = \emptyset$}
    \STATE mark $s_j$ unwritable; \textbf{continue}
  \ENDIF
  \STATE move $c$ to $\arg\max_{p \in F} \textsc{Affinity}(c,p)$, and rewrite every
         reference to $c$ (\S\ref{sec:deploy})
\ENDFOR
\RETURN marked repository
\end{algorithmic}
\end{algorithm}

\begin{algorithm}[t]
\caption{\sys{} extraction}
\label{alg:extract}
\begin{algorithmic}[1]
\STATE \textbf{input:} repository $R'$, key $k$, payload length $\ell > 0$, width $E \ge 1$,
       repetition $r$ odd and positive
\STATE $C \leftarrow \textsc{Eligible}(R')$
\STATE $U \leftarrow \textsc{AdmissibleDestinations}(R')$
\STATE \textbf{abort} if $|U| < 2$
\STATE $n \leftarrow \lfloor \log_2 |U| \rfloor$; \ $\varrho \leftarrow 2^{n}$
\STATE \textbf{abort} if $E > n$ \textbf{or} $\lceil \ell r / E \rceil > |C|$
\STATE $O \leftarrow \textsc{Sort}\big(C, \ \text{key}{=}\mathrm{PRF}_k(\text{``order''}, \mathrm{mid}(\cdot))\big)$
\STATE $b \leftarrow \langle \, \rangle$
\FOR{$j \leftarrow 0$ \textbf{to} $\lceil \ell r / E \rceil - 1$}
  \STATE $c \leftarrow O[j]$
  \STATE $a \leftarrow \textsc{Sort}\big(U, \text{key}{=}\mathrm{PRF}_k(\text{``cand''}, \mathrm{mid}(c), \cdot)\big)[0{:}\varrho]$
  \STATE $i \leftarrow \textsc{IndexOf}(a, \mathrm{pkg}(c))$ \COMMENT{$\bot$ if outside the shortlist}
  \STATE append the $E$ bits of $(i \bmod 2^{E})$ to $b$, or $0^{E}$ if $i = \bot$
\ENDFOR
\RETURN \textsc{MajorityVote}$(b[0{:}\ell r], \ell, r)$ \COMMENT{padding discarded}
\end{algorithmic}
\end{algorithm}

\subsection{Carrier selection}
\label{sec:carriers}

The decoder sees only the marked tree, so it must recompute the carrier set and the reading order
and obtain exactly what the encoder obtained, even after an adversary has moved an arbitrary subset
of modules. Writing $S(R)$ for either quantity derived from repository $R$, the scheme is correct
only if
\begin{equation}
S(R) = S(\rho(R)) \quad \text{for every relocation } \rho \text{ preserving } \mathcal{I} ,
\label{eq:invariance}
\end{equation}
where $\mathcal{I}$ is the set of inputs the two quantities are computed from: each module's file
name and top-level definition names, the eligibility tests of this section, and the destination
universe. A relocation preserves $\mathcal{I}$ when it moves module files between packages of the
tree, which is the transformation the scheme is designed to survive; moving a module to the top
level or out of the package does not, and removes it from the carrier set.

This is more restrictive than it appears, because it rules out any test that consults a module's
current path: relocation is exactly what changes that path. The cost of violating it is not
proportional to the mistake. Carriers are read in a keyed order, so a single carrier gained or lost
displaces every symbol after it, and a rule that misclassifies a handful of modules can corrupt most
of the payload.

The rules below therefore read a module's own file name and top-level definitions rather than its
location, with one implementation exception noted at the end of this section. Eq.~\eqref{eq:invariance} is to be read with that scope: the carrier set and the reading
order are stable under any transformation that leaves the identity inputs, the eligibility inputs
and the destination universe unchanged, which is what relocation does. It is not a guarantee for
every conceivable edit. Moving a module to the top level, for instance, removes it from the carrier
set, because a module with no enclosing package has no placement to encode.

\textbf{Location-free module identity.} A carrier is identified by
\begin{equation}
\mathrm{mid}(m) = H\!\left(\text{leaf}(m) \,\|\, \mathrm{sort}(\text{defs}(m))\right),
\end{equation}
the SHA-256 hash of the module's file name together with the sorted names of its top-level
definitions. Neither input mentions a package, so the decoder recognises the module wherever it now
sits. Module-level imports are deliberately excluded: the encoder rewrites imports when it moves a
module, so an import-sensitive identity would change during embedding and desynchronise the decoder
from the encoder that produced the tree.

\textbf{Which modules may carry.} A module is a candidate when it is addressable and free to move.
Addressable means its file name and its identity are both unique in the tree, so that the decoder
can name it and no destination can collide with an existing file; package initialisers, dunder
modules, and modules already at the top level are excluded because they have no package to vary or
because the import machinery depends on where they sit.

Free to move means nothing outside the import graph depends on the module's location. Two things
can: a framework that finds the module by enumerating a directory, such as a migration or a
management command, and an addressing scheme that names the module rather than importing it, whether
from a configuration file, from a dotted path in the module's own text, or from the module reading
\texttt{\_\_file\_\_} to reach a neighbouring file.

Identity uses the module's own content alone. Eligibility, as a general rule, is decided from the module's syntax tree
and file name together with the project's configuration files, which relocation does not move and
which name modules by a leaf that relocation does not rename, because only a location-free
formulation satisfies Eq.~\eqref{eq:invariance}; one exception in the implementation we measured is
given below. \S\ref{sec:eval:invariance} measures what the path-based alternative costs.

One exclusion in the implementation we measured is an exception and should be stated. A module whose
dotted path begins with \texttt{checkers} is excluded by that path, a rule predating the syntactic
formulation and retained because \textsc{Pylint} discovers its checkers by walking that directory.
It matches $51$ modules on \textsc{Pylint} and none on the other three repositories, so the measured
configuration is location-free on \textsc{Django}, \textsc{Airflow} and \textsc{Salt} but not on
\textsc{Pylint}, where Eq.~\eqref{eq:invariance} is consequently not guaranteed under relocations
that cross that directory. \S\ref{sec:eval:carrieraware} measures what that costs. The exclusions are not free, and the cost is uneven: of the $211$ \textsc{Django} modules
that pass the addressability tests, the discovery-bound rule removes $51$, a quarter of them,
because \textsc{Django} has migration, management-command and template-tag directories. It removes none on
\textsc{Airflow} or \textsc{Salt}, which have no such directories; on \textsc{Pylint} the
corresponding hazard is handled by the \texttt{checkers} rule given below. The
\texttt{\_\_file\_\_} exclusion is a default-off switch, not used in the configuration measured
here; enabling it would remove a further $0.3$ to $4.1\%$.

\subsection{Watermark embedding}
\label{sec:embedding}

\textbf{Admissible destinations.} The destination universe $U$ is obtained by enumerating the
packages of the tree and removing those that are unsafe to land in. Importing a submodule executes
every ancestor package initialiser, so a package whose initialiser chain pulls in an optional
third-party dependency is unsafe: a module that lands there acquires that dependency, and every
importer of it then fails when the dependency is absent. A package that a framework enumerates by
name is unsafe for the mirror-image reason, since a module that lands there acquires a meaning it
never had.

\textbf{Nominal capacity.} With $\varrho$ candidate destinations per carrier and $|C|$ carriers, the
configuration set has $\varrho^{|C|}$ nominal combinations before joint feasibility is imposed, so
the channel holds at most
\begin{equation}
|C| \cdot \log_2 \varrho \ \text{bits}.
\label{eq:capacity}
\end{equation}
We call this the \emph{nominal capacity}, and it is an upper bound throughout this paper rather
than a payload the encoder is guaranteed to write. It counts every carrier as freely placeable and
treats the choices as independent, whereas import cycles and name collisions couple them and can
leave a destination unreachable for a given module. \S\ref{sec:eval:wet} measures the gap. \S\ref{sec:eval:capacity} measures it on ten repositories and \S\ref{sec:eval:effectiveness}
measures how much of it is usable.

\textbf{Keyed candidate shortlist.} For carrier $c$, the key induces a pseudorandom ordering of $U$
through $\mathrm{PRF}_k(\text{``cand''}, \mathrm{mid}(c), \cdot)$, truncated to the first $\varrho$
packages; this is $c$'s candidate list $a(c)$. If $c$ occupies the $i$-th candidate, counting from
zero, it carries the symbol $i \bmod 2^{E}$. The radix is the largest power of two at most $|U|$, so that the $2^{E}$ symbol values partition
the candidate list into equivalence classes of equal size. Equal classes require only that $2^{E}$ divide the radix; taking the largest power of two at most
$|U|$ guarantees this for every $E$, which is why we adopt it, not because no other radix could
work.

\textbf{Symbol radix and encoder slack.} Choosing $E$ strictly below $\log_2 \varrho$ is deliberate.
It leaves $\varrho / 2^{E}$ distinct packages encoding each symbol value, so the encoder can satisfy
a required symbol in several ways and pick the most defensible destination, and a module that cannot
be moved does not by itself make the message unwritable. Setting $E$ at its ceiling collapses this
freedom to one and leaves an immovable carrier no alternative destination.

\textbf{Keyed reading order.} Carriers are read in the order induced by
$\mathrm{PRF}_k(\text{``order''}, \mathrm{mid}(\cdot))$, not in alphabetical or tree order. A message
of $b$ bits occupies the first $\lceil b/E \rceil$ carriers in that order and no others. On
\textsc{Django}, $32$ of $160$ eligible carriers hold a $64$-bit payload, and which $32$ is a
function of the key. An adversary who knows the scheme completely can therefore compute the carrier
set but not the payload-bearing subset, so any budget it spends is spread across carriers of which
most are irrelevant.

\textbf{Destination choice.} Within the equivalence class $K$ the encoder is free, and it resolves
that freedom by a vocabulary-overlap score. The module's file name and path are split into
subtokens, each candidate package is described by the subtokens of the modules it already contains,
and the encoder selects the feasible package with the largest overlap. We call this score
\textsc{Affinity}, as in Algorithm~\ref{alg:embed}. The decoder never learns which destination was
chosen, or why; it reads only $i \bmod 2^{E}$. This choice therefore sits outside the codec and can be replaced without affecting
correctness.

\textbf{Two encoder variants.} Algorithm~\ref{alg:embed} as written is the \emph{raw} encoder: it
chooses a destination per carrier and moves. The variant that ships wraps it in an
\emph{import-verifying} loop, which re-checks the import graph after each round of moves, excludes
any placement that broke it, and retries. The two differ only in which placements they are willing
to use, never in how a symbol is read, so the decoder is the same for both. Every measurement in
this paper uses the import-verifying encoder except Table~\ref{tab:wet}, which isolates the channel
from the loop and says so.

\textbf{Repetition coding.} Each payload bit may be stored $r$ times. The codeword is the payload
bit string concatenated with itself $r$ times, and extraction takes a majority vote across the $r$
copies of each bit position, correcting up to $\lfloor (r{-}1)/2 \rfloor$ corrupted copies. No
separate interleaver is needed: because the reading order is a keyed permutation of the carriers, the
$r$ copies of one payload bit are assigned to carriers in unrelated parts of the tree. Repetition
costs carriers rather than code, and the price is paid in capacity, which Eq.~\eqref{eq:capacity}
makes explicit.

\subsection{A worked example}
\label{sec:example}

Table~\ref{tab:example} embeds six bits in a repository of three carriers. The destination universe
holds eight admissible packages, so the radix is $\varrho = 8$ and each carrier offers three bits; we
set $E = 2$, which leaves two packages per symbol value. The payload $10\,01\,11$ becomes the symbol
sequence $s = \langle 2, 1, 3 \rangle$.

The carriers are read in the keyed order $c_2, c_3, c_1$, not in alphabetical or tree order, and each
gets its own keyed shuffle of the universe. Carrier $c_2$ must carry $s_0 = 2$, which is encoded by
the two shortlist positions $i \in \{2, 6\}$; it already sits in \texttt{cache} at position $2$, so
it does not move. Carrier $c_3$ must carry $1$ and sits outside $\{\texttt{api}, \texttt{db}\}$, so
it moves to \texttt{api}. Carrier $c_1$ moves to \texttt{utils}. Two modules move and one stays.

The decoder is given the key and the payload length. It recomputes the carrier set, the universe,
the radix, and the reading order, looks up each carrier's current package in its own shortlist, and
reads $2, 1, 3$ off the indices $2, 1, 7$ modulo $4$.

Now suppose an adversary relocates $c_3$ from \texttt{api} to \texttt{text}, position $4$ in its
shortlist. The decoder reads $4 \bmod 4 = 0$ instead of $1$, so the symbol $01$ is read as $00$ and
one payload bit is lost. It loses nothing else: $c_3$'s identity is computed from its file name and
its definitions, so the relocated module is still recognised as a carrier and still occupies the
second position in the reading order. Had eligibility been decided from the directory a module
currently sits in, the relocated module could have left the carrier set entirely, $c_1$ would have
shifted from the third position to the second, and every symbol after the attacked one would have
been read from the wrong carrier. This is the distinction \S\ref{sec:carriers} is built around and
\S\ref{sec:eval:invariance} measures.

\begin{table}[t]
\caption{Embedding $10\,01\,11$ in three carriers with $|U| = 8$, $\varrho = 8$, $E = 2$. Each row
is one carrier's keyed candidate shortlist. The symbol is the shortlist index modulo $2^{E} = 4$, so
two positions per row encode the required value; the chosen one is in bold. $^{\dagger}$Carrier
$c_2$ is already in an encoding package and does not move. This example is constructed, not
measured.}
\label{tab:example}
\centering
\scriptsize
\setlength{\tabcolsep}{2.6pt}
\begin{tabular}{clcccccccccl}
\toprule
 & & \multicolumn{8}{c}{candidate shortlist $a(c)$, by index $i$} & & \\
\cmidrule(lr){3-10}
$j$ & $c$ & 0 & 1 & 2 & 3 & 4 & 5 & 6 & 7 & $s_j$ & placed \\
\midrule
0 & $c_2$ & db & utils & \textbf{cache} & core & api & text & io & net & 2 & cache$^{\dagger}$ \\
1 & $c_3$ & net & \textbf{api} & core & io & text & db & utils & cache & 1 & api \\
2 & $c_1$ & text & cache & io & api & db & core & net & \textbf{utils} & 3 & utils \\
\bottomrule
\end{tabular}
\end{table}

\subsection{Watermark extraction}

\textbf{Carrier identification.} Extraction begins by re-running carrier selection on the repository
in hand. Because those rules read module content rather than location, the result is the encoder's
carrier set after any rearrangement that leaves $\mathcal{I}$ intact. \S\ref{sec:eval:invariance}
tests this directly, by
relocating half of a repository and comparing the two sets.

\textbf{Extraction from an unmodified repository.} Every carrier the encoder moved sits in a package
drawn from its own shortlist, so its index is defined and the symbol is read back exactly. A carrier
the encoder left in place need not: if its home package is outside its shortlist the index is
undefined and the decoder emits zeros, and if it is inside, the position carries whatever symbol
that package encodes. Either way the decoder cannot tell such a position from a written one and
reads it as it stands. Recovery from an unmodified
repository is therefore exact when the selected payload positions were all writable. No reference
copy, trained model, or stored record of where the modules were placed takes part.

\textbf{Extraction from a modified repository.} A relocated carrier may sit outside its shortlist, in
which case its index is undefined. \sys{} then emits zeros for that symbol rather than failing, so a
damaged repository yields a partially recovered payload instead of nothing. Degradation is graded in
the number of carriers the adversary actually disturbed, and the repetition code absorbs the first
several.

\subsection{Reference rewriting}
\label{sec:deploy}

An embedding rewrites the references to a module it moves, so that the marked repository resolves as
the original did. Three forms are rewritten: import statements, anywhere in the repository; string
literals holding a dotted module path, which frameworks use to address modules by name in settings
and entry-point tables; and attribute access through a parent package, as in
\texttt{generic.list.ListView}, which carries no import statement and is resolved from the file's
own syntax tree. These are the forms a static pass can find. A reference assembled at run time, from
a configuration value or by string concatenation, is not among them, which is why eligibility
excludes modules addressed that way (\S\ref{sec:carriers}) rather than relying on the rewriter to
repair them. Every rewritten file is compiled before it is written, and configuration
files are never written, since a module's eligibility depends on not being named from one.

These rewrites insert no code and delete none, and none moves a module, so none can change a
placement. Whether they change what the decoder recomputes is a separate question, and we measured
it: on each of the four repositories we snapshotted every carrier's identity and package together
with the destination universe, ran both rewriters with the encoder's own moves, and re-read them.
Both are identical afterwards as sets, not merely in size, at $160$ carriers over a universe of $32$
on \textsc{Django}, $361$ over $65$ on \textsc{Airflow}, $412$ over $33$ on \textsc{Salt} and $83$
over $10$ on \textsc{Pylint}, with the payload still extracting exactly.

\section{Evaluation}
\label{sec:eval}

\subsection{Experimental setup}
\label{sec:setup}

\textbf{Repositories.} We evaluate on four production Python repositories, each pinned at a fixed
commit: \textsc{Django} (\texttt{97aa3b7f08f5}, $864$ modules), \textsc{Airflow}
(\texttt{3adbbe1c58e4}, $698$), \textsc{Salt} (\texttt{25f141f0ccc9}, $967$) and \textsc{Pylint}
(\texttt{5f7e4013c004}, $180$). They span roughly five-fold in size and, more importantly for our
purposes, differ in architectural style, from \textsc{Airflow}'s wide package tree to
\textsc{Pylint}'s flat one. Results are reported per repository throughout and never pooled.

One property of \textsc{Airflow} bears on which experiments it appears in. Its package is not a
self-contained directory: fourteen of its subdirectories are symbolic links into the surrounding
monorepo. Measured where it stands, with those links resolving, it has $698$ modules and $361$
carriers, and it carries and returns a payload exactly. An experiment that copies the tree must
preserve those links to measure the same repository; the figures reported here do. This is a
property of that checkout, not of the scheme.

\textbf{Parameters.} Unless stated otherwise the payload is $64$ bits, $E$ is chosen per repository
to hold the encoder's freedom constant at eight packages per symbol ($E=2$ for \textsc{Django} and
\textsc{Salt}, $3$ for \textsc{Airflow}, $1$ for \textsc{Pylint}, whose universe caps it at four),
and no error correction is applied, so $r = 1$ and each payload bit is stored exactly once. Every
robustness figure is therefore a floor rather than a limit, and
Table~\ref{tab:relocation} measures what a code recovers. Each table states its own coding configuration and the keys and attack seeds it
averages over.

\textbf{Reporting convention.} Recovery is reported as the bit error rate (BER) of the extracted
payload, on the scale $0$ to $1$: $\mathrm{BER}=0$ means every payload bit was recovered,
and for an unbiased binary payload $\mathrm{BER} \approx 0.5$ is chance-level accuracy under this
decoder, which we report as such rather than as a statement about the information the artifact still
holds. Where it aids interpretation we also give the absolute count of wrong bits out of the payload
length.

\textbf{Comparability.} The relocation, payload-size and carrier-aware grids are separate
experiments with their own key sets and attack seeds, so the same nominal condition can differ
between them: untargeted relocation of $5\%$ of \textsc{Django} reads $0.017$, $0.034$ and $0.048$
in Tables~\ref{tab:relocation}, \ref{tab:sweep} and \ref{tab:carrieraware} respectively. Comparisons are matched
within a table, never across them, and dispersion is reported where arms are compared.

\textbf{Protocol.} Every grid is executed twice into separate files and compared field by field.

\subsection{Capacity}
\label{sec:eval:capacity}

Table~\ref{tab:capacity} reports the channel of each repository. Nominal capacity is the product of
the number of eligible carriers and the bits each carries, which is an upper bound on what the
encoder can write.

\begin{table}[t]
\caption{Nominal capacity per repository, an upper bound. $|U|$ is the number of admissible
destination packages and $\varrho$ the radix. The four repositories above the rule are those used in the
experiments, each table naming its own subset; the six below extend the capacity measurement
only.}
\label{tab:capacity}
\centering
\begin{tabular}{lrrrrr}
\toprule
Repository & Modules & Carriers & $|U|$ & $\varrho$ & Capacity \\
\midrule
\textsc{Django}   & 864 & 160 & 32 & 32 & 800\,b \\
\textsc{Airflow}  & 698 & 361 & 65 & 64 & 2{,}166\,b \\
\textsc{Salt}     & 967 & 412 & 33 & 32 & 2{,}060\,b \\
\textsc{Pylint}   & 180 &  83 & 10 &  8 & 249\,b \\
\midrule
\textsc{Ansible}  & 576 & 241 & 46 & 32 & 1{,}205\,b \\
\textsc{Home Assistant} & 9{,}847 & 239 & 782 & 512 & 2{,}151\,b \\
\textsc{SymPy}    & 854 & 300 &  7 &  4 & 600\,b \\
\textsc{scikit-learn} & 335 & 170 & 10 & 8 & 510\,b \\
\textsc{Odoo}     & 268 & 121 &  8 &  8 & 363\,b \\
\textsc{idlelib} (CPython) & 126 & 63 & 1 & 1 & \textbf{0\,b} \\
\bottomrule
\end{tabular}
\end{table}

Two observations follow. First, capacity is set by architecture rather than by size.
\textsc{Airflow} holds more than twice \textsc{Django}'s payload from fewer modules, because its
wider package tree offers more legal destinations per module. \textsc{Home Assistant}, an order of
magnitude larger than any other repository here, holds no more than \textsc{Airflow}, because its
thousands of integration modules share a few dozen file names and a carrier must be uniquely named.

Second, the precondition is real: \textsc{CPython}'s \texttt{idlelib} has a single admissible
destination and therefore no capacity at all. A repository organised as one flat package, or one
whose every package initialiser pulls in an optional dependency, is outside the scheme's scope. This
is the analogue of the minimum-function-count condition that binary relocation watermarks
state~\cite{softmark}, and it should be checked before the scheme is applied.

\subsection{Effectiveness and cost}
\label{sec:eval:effectiveness}

Table~\ref{tab:cost} reports what embedding a $64$-bit payload costs. The encoder moves between $26$
and $44$ modules and rewrites references in between $17$ and $82$ files, adding and deleting none.

\begin{table}[t]
\caption{Cost of embedding a $64$-bit payload, uncoded ($r=1$), mean over three keys. ``Edited''
counts files whose references the encoder rewrote; none is added or deleted.}
\label{tab:cost}
\centering
\begin{tabular}{lrrrr}
\toprule
Repository & Moved & Edited & Diff lines & Bits/move \\
\midrule
\textsc{Django}  & 29.7 & 81.3 & 442.3 & 2.16 \\
\textsc{Salt}    & 26.0 & 17.3 &  75.3 & 2.47 \\
\textsc{Pylint}  & 43.7 & 54.3 & 315.0 & 1.47 \\
\bottomrule
\end{tabular}
\end{table}

Embedding efficiency, the payload bits delivered per module moved, ranges from $1.47$ on
\textsc{Pylint} to $2.47$ on \textsc{Salt}, and tracks the symbol width $E$: a larger $E$
carries more bits per carrier and therefore needs fewer moves.

Table~\ref{tab:utilisation} reports recovery on an unattacked tree as the carrier budget is
consumed. Utilisation here is the share of available carriers the codeword occupies, not the payload
as a share of nominal capacity.

\begin{table}[t]
\caption{Recovery against utilisation on \textsc{Django}, uncoded ($r=1$). Utilisation is
$\lceil b/E \rceil$ over the $160$ carriers available. Wrong bits on an unattacked tree, per key.
Each key here carries a payload derived from that key, where Tables~\ref{tab:transforms}
to~\ref{tab:relocation} carry one fixed payload under the same three keys.}
\label{tab:utilisation}
\centering
\begin{tabular}{lrrccc}
\toprule
Repository & Payload & Util. & key1 & key2 & key3 \\
\midrule
\multirow{4}{*}{\textsc{Django}}
 &  32\,b & 10\% & 0 & 0 & 2 \\
 &  64\,b & 20\% & 0 & 0 & 2 \\
 & 128\,b & 40\% & 0 & 0 & 2 \\
 & 256\,b & 80\% & 3 & 3 & 3 \\
\bottomrule
\end{tabular}
\end{table}

Two of the three keys recover exactly up to $40\%$ utilisation; the third loses two bits at every
payload size, and at $80\%$ every key loses three. A position the encoder cannot write stays
unwritten, and because the decoder cannot tell such a position from a written one it reads it as an
ordinary symbol. This is the classical non-shared selection channel~\cite{fridrich2005wet}, for
which codes exist that recover the loss without telling the decoder which positions were skipped;
we do not use one, so the figures here are a floor. \S\ref{sec:eval:wet} evaluates a different
encoder configuration over thirty keys and bears on the residual error seen here.

\subsection{Robustness to code transformation}
\label{sec:eval:transforms}

The transformations that published code watermarks are evaluated against rewrite the text of source
files. Table~\ref{tab:transforms} reports what they cost \sys{}. Every one of them costs exactly zero
bits, with zero variance, on every repository and trial we ran. The payload is carried by placement rather than
by source text, and each of these transformations preserves the two features carrier identity is
built from, the file name and the set of top-level definition names, so both the channel and the
carrier set are left as the encoder produced them. Transformations that do change those features,
such as renaming files or renaming top-level definitions, are a different case and were not part of
this menu.

\begin{table}[t]
\caption{Content transformations run against \sys{}, uncoded ($r=1$), $64$-bit payload.
``Files'' is the range over \textsc{Django}, \textsc{Salt} and \textsc{Pylint}. The lower block is
the removal menu reported for content-carrier watermarks, at nine trials per cell per repository,
$216$ in total, applied in sequence for ``Combined''. The two blocks are separate experiments.}
\label{tab:transforms}
\centering
\setlength{\tabcolsep}{3.0pt}\small
\resizebox{\columnwidth}{!}{%
\begin{tabular}{llcc}
\toprule
Transformation & Rewrites & Files & BER \\
\midrule
Reformatting         & whitespace, quotes, breaks & 9--478   & \textbf{0.0000} \\
Import sorting       & import order, grouping    & 116--443 & \textbf{0.0000} \\
Comment removal      & every comment             & 180--958 & \textbf{0.0000} \\
\midrule
Rename 25\% locals    & local variables           & 180--958 & \textbf{0.0000} \\
Rename 50\% locals    & local variables           & 180--958 & \textbf{0.0000} \\
Rename 75\% locals    & local variables           & 180--958 & \textbf{0.0000} \\
Rename 100\% locals   & local variables           & 180--958 & \textbf{0.0000} \\
Dead-code insertion   & unreachable blocks        & 180--958 & \textbf{0.0000} \\
Exception wrapping    & wrapped bodies            & 180--958 & \textbf{0.0000} \\
Print insertion       & inserted prints           & 180--958 & \textbf{0.0000} \\
Combined              & \begin{tabular}[t]{@{}l@{}}100\% renaming, dead code,\\
                        exception wrapping, prints\end{tabular} & 510--2{,}786 & \textbf{0.0000} \\
\bottomrule
\end{tabular}}
\end{table}

The lower block of Table~\ref{tab:transforms} is the removal menu those papers report, run against
marked repositories and measured as the same quantity they report. Across the $216$ measured trials
of this menu, every one recovered the payload exactly. The combined attack totals $1{,}827$ file rewrites on
\textsc{Django} and $2{,}786$ on \textsc{Salt}, where a file touched by three of the four
transformations is counted three times.

Relocation, the attack that does damage \sys{}
(\S\ref{sec:eval:relocation}), is not defined for a function-level carrier and appears in no
published attack menu.\footnote{One observation from running these attacks at
repository scale is not visible when they are evaluated on isolated functions. Renaming a function
\emph{parameter} is semantics-preserving within one function but not within a project, because
callers pass parameters by keyword and a repository-wide parameter rename changes the public call
signature. The attack we report therefore renames local variable names and leaves parameters
intact, which is the strongest form of it that yields a working artifact.}

\subsection{Comparison with content-carrier watermarks}
\label{sec:eval:comparison}

The systems closest to ours in purpose hide the payload in the code text. A numerical head-to-head
is not available: their datasets are collections of standalone functions, which have no directory
structure and therefore no channel for us to write to, and the published figures use different
corpora, different attack implementations and, in places, different metrics. We therefore compare
what the carriers assume and what each attack surface exposes, and quote individual published
figures only with their setting attached.

\begin{table}[t]
\caption{What the carriers assume and what they expose. ``Blind'' means extraction needs the
artifact, the key and the agreed decoding parameters, with no original copy, trained extractor or
stored placement record.}
\label{tab:headtohead}
\centering
\setlength{\tabcolsep}{3pt}\footnotesize
\begin{tabular}{@{}p{14mm}p{30mm}p{30mm}@{}}
\toprule
 & Content-carrier schemes & \sys{} \\
\midrule
Carrier        & tokens, identifiers, statements & which package a module occupies \\[2pt]
Unit           & a function & a repository \\[2pt]
Capacity grows & with the number of functions & with carriers $\times \log_2 \varrho$ \\[2pt]
Exposed to     & renaming, rewriting, regeneration & relocation; renaming of files or top-level
                 definitions \\[2pt]
Untouched by   & file relocation leaves the embedded features unchanged; extraction under
                 relocation untested & reformatting, comment removal, local renaming \\[2pt]
Blind          & scheme-dependent & yes \\
\bottomrule
\end{tabular}
\end{table}

Two published figures give the scale of the exposure each carrier has. The Transformer variant of
\textsc{SrcMarker} reports $97.26\%$ bit accuracy with no attack and $62.68\%$ after full variable
substitution on CSN-Java~\cite{srcmarker}; \textsc{CLASP} reports $98.05\%$ and $76.93\%$ for its own
no-attack and full-renaming settings, which are measured on different datasets and are not a single
degradation curve~\cite{clasp}. Against the same class of transformation \sys{} loses no bit in any
of the $216$ trials of \S\ref{sec:eval:transforms}, because those transformations do not move a file
or rename one. The exposure runs the other way for relocation, which we measure in
\S\ref{sec:eval:relocation} and which a content carrier would not be expected to notice, though we
have not run a content-carrier scheme under relocation and do not report a figure for it.

The two carriers therefore have complementary attack surfaces rather than disjoint ones: a
refactoring that renames files while reorganising them touches both. Where a per-function scheme can go further is capacity, since its payload accumulates with the
number of functions while ours is bounded by carriers and radix, at the at most $2{,}166$ bits
measured here. The two have not been measured on a common corpus, so this is a property of the
constructions rather than a ranking.

\subsection{Robustness to relocation}
\label{sec:eval:relocation}

Relocation is the attack a placement carrier is exposed to by construction. We model an adversary
without the key that moves a fraction of \emph{all} modules, not merely carriers, into other legal
packages, so its budget is spread over the repository rather than concentrated. Because the
eligibility rule is public under our threat model, a keyless adversary could instead compute the
eligible carrier set and spend its budget there; it would still not know which carriers hold the
payload, nor the order in which they are read. This section reports the untargeted strategy;
\S\ref{sec:eval:carrieraware} reports the carrier-aware one and compares the two at an equal number
of moves at the $5\%$ and $20\%$ budgets. Table~\ref{tab:relocation} reports the bit error rate the untargeted adversary produces.

\begin{table}[t]
\caption{Relocation robustness on \textsc{Django} at a $64$-bit message, over three keys and three
attack seeds, uncoded and under triple repetition at equal message length. BER runs from $0$ (every
bit recovered) to $1$, with $0.5$ the chance level; wrong message bits out of $64$ in parentheses.}
\label{tab:relocation}
\centering
\begin{tabular}{llcccc}
\toprule
Code & Rate & no attack & 5\% & 20\% & 50\% \\
\midrule
Uncoded ($r=1$)       & 1   & 0.0000 & 0.0174 & 0.1076 & 0.2483 \\
                      &     & (0.0)  & (1.1)  & (6.9)  & (15.9) \\
\midrule
Repetition ($r=3$)    & 1/3 & \textbf{0.0000} & \textbf{0.0000} & \textbf{0.0330} & \textbf{0.1528} \\
                      &     & (0.0)  & (0.0)  & (2.1)  & (9.8) \\
\bottomrule
\end{tabular}
\end{table}

\begin{figure*}[t]
\centering
\setlength{\tabcolsep}{4pt}
\renewcommand{\arraystretch}{0.9}
\begin{tabular}{cccccc}
\footnotesize original & \footnotesize $5\%$ & \footnotesize $10\%$ &
\footnotesize $20\%$ & \footnotesize $30\%$ & \footnotesize $50\%$ \\
\includegraphics[width=0.148\textwidth]{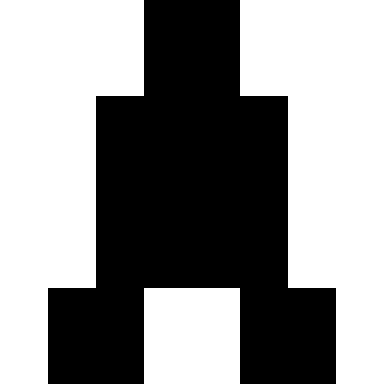} &
\includegraphics[width=0.148\textwidth]{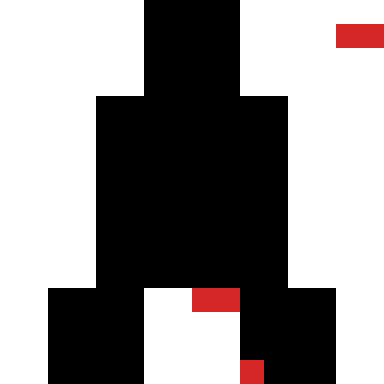} &
\includegraphics[width=0.148\textwidth]{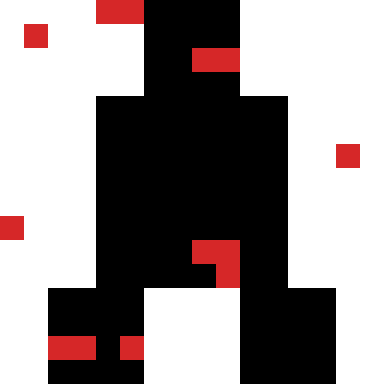} &
\includegraphics[width=0.148\textwidth]{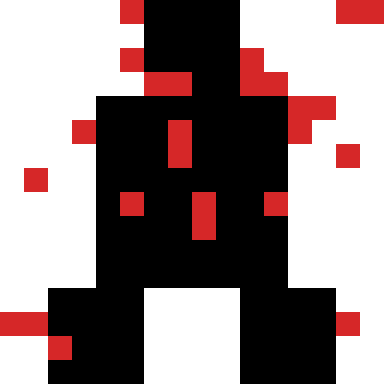} &
\includegraphics[width=0.148\textwidth]{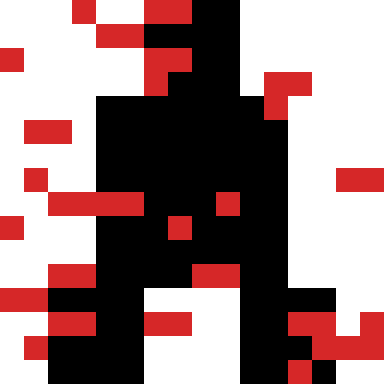} &
\includegraphics[width=0.148\textwidth]{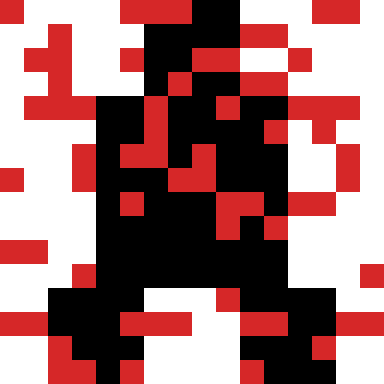} \\
\footnotesize 0 of 256 & \footnotesize 5 & \footnotesize 13 &
\footnotesize 25 & \footnotesize 42 & \footnotesize 67 \\
\end{tabular}
\caption{The payload recovered from \textsc{Salt} as relocation increases. The payload is a
$16 \times 16$ binary glyph carried as $256$ raw bits, so a wrong bit is a wrong pixel and the
recovered message can be looked at rather than inferred from a rate. Pixels that differ from the
original are filled red and counted below each panel. Uncoded ($r=1$), $E=2$, three keys and three
attack seeds per level; the panel shown is the trial whose bit error is closest to the mean of its
level, and those means are $0.021$, $0.053$, $0.094$, $0.164$ and $0.267$. With nothing relocated
the recovered glyph is identical to the original in all three keys, which is why no panel for that
level is shown separately.}
\label{fig:glyph}
\end{figure*}

Degradation is graded rather than abrupt where the carrier set survives the attack, which on
\textsc{Pylint} it does not (\S\ref{sec:eval:carrieraware}). At half the repository relocated the payload retains substantial
information: on \textsc{Django} with $r = 3$, about $10$ of $64$ bits are wrong against a chance
floor of $32$, and there is no threshold below which the mark is intact and above which it is gone.
Fig.~\ref{fig:glyph} shows the same progression directly on \textsc{Salt}, at a $256$-bit payload
carried as an image so that the recovered message can be read off. The count of wrong pixels rises
from $5$ to $67$ of $256$ as relocation goes from $5\%$ to $50\%$, and the glyph stays recognisable
throughout. The carrier set is unchanged in every row of this grid, constant at $160$ modules on
\textsc{Django} and $412$ on \textsc{Salt}, which is the direct empirical test of
Eq.~\eqref{eq:invariance}; \S\ref{sec:eval:carrieraware} repeats it on \textsc{Salt}.

\textbf{What a code buys.} Table~\ref{tab:relocation} also reports triple repetition at equal
message length. It removes the residual error at $5\%$ relocation and cuts the error at $50\%$ by
nearly $40\%$, which costs capacity rather than code: the keyed reading order already scatters the
three copies of a bit across unrelated positions in the carrier order, so no interleaver is needed,
though nothing prevents two copies sharing a package. The uncoded figures reported elsewhere in this
paper are therefore a floor.

\textbf{Payload size.} Table~\ref{tab:sweep} sweeps the payload from $32$ to $256$ bits. Two effects
are visible and they pull in opposite directions. The error \emph{rate} is higher at the larger payloads in these runs: at $50\%$ relocation
\textsc{Django} loses $22\%$ of a $32$-bit payload and $30\%$ of a $256$-bit one. A larger payload
occupies more carriers and so exposes more of them, which accounts for the rise in the absolute
count; we report the rate as observed rather than derive it.
The absolute number of surviving bits nonetheless rises steeply, from $25$ correct bits at $32$~b to
$180$ at $256$~b. Which of these matters depends on the application and on the verification criterion, neither of
which we evaluate here: exact recovery and partial recognisability are different tests, and the
choice also interacts with the error-correction rate.

\begin{table}[t]
\caption{Payload size against relocation, uncoded ($r=1$), over two attack seeds. Three keys for
\textsc{Django}. Wrong bits / payload bits. A different attack-seed realisation from
Table~\ref{tab:relocation}.}
\label{tab:sweep}
\centering
\begin{tabular}{lrccc}
\toprule
Repository & Payload & 5\% & 20\% & 50\% \\
\midrule
\multirow{4}{*}{\textsc{Django}}
 &  32\,b & 2.2/32 & 3.3/32 & 7.2/32 \\
 &  64\,b & 2.2/64 & 7.5/64 & 17.5/64 \\
 & 128\,b & 3.2/128 & 21.5/128 & 37.7/128 \\
 & 256\,b & 9.0/256 & 33.5/256 & 75.7/256 \\
\bottomrule
\end{tabular}
\end{table}

\textbf{Is this a realistic attack budget?} The relocation levels above span a range; they do not
claim to describe what happens to software. To calibrate them we measured how much real repositories
rearrange themselves, by diffing the trees of their own release tags, over six projects with $115$ to
$744$ tags each spanning $14$ to $21$ years, and counting the \texttt{.py} files whose directory
changes.

\begin{table}[t]
\caption{Relocation in the wild: share of \texttt{.py} files that change directory, measured between
a repository's own release tags.}
\label{tab:churn}
\centering
\begin{tabular}{lrrrr}
\toprule
& \multirow{2}{*}{Tags} & \multicolumn{2}{c}{Per release} & Per year \\
\cmidrule(lr){3-4}
Repository & & median & max & median \\
\midrule
\textsc{Django}       & 519 & 0.14\% & 46.4\% & 0.26\% \\
\textsc{Pylint}       & 223 & 0.00\% & 98.3\% & 1.39\% \\
\textsc{Ansible}      & 744 & 0.14\% & 56.0\% & 0.67\% \\
\textsc{Salt}         & 331 & 0.04\% &  3.9\% & 0.19\% \\
\textsc{scikit-learn} & 155 & 0.00\% & 63.0\% & 0.52\% \\
\textsc{SymPy}        & 115 & 0.00\% &  2.4\% & 0.52\% \\
\midrule
\textbf{Median of the above} & & \textbf{0.02\%} & & \textbf{0.52\%} \\
\bottomrule
\end{tabular}
\end{table}

Two things stand out. Ordinary releases move almost nothing: the median release of every project
moves at most $0.14\%$ of its \texttt{.py} files between directories, and a median year moves about
half a percent. Against that baseline our mildest attack, at $5\%$, moves an order of magnitude more
than a median year of maintenance.

The maximum column is very different. Four of the six projects have undergone at least one release
in which a large fraction of the tree was rearranged at once, up to $98\%$ for \textsc{Pylint},
because projects are occasionally restructured deliberately. These are counts of files that changed
directory, and nothing more: we did not extract a payload across any of these releases, so they
calibrate the attack budgets against real development rather than establish what the mark survives.
A release also renames files and definitions, which the counts here do not capture and which
\S\ref{sec:eval:identity} shows matters more.

\subsection{Selection-channel invariance, measured}
\label{sec:eval:invariance}

Section~\ref{sec:carriers} claims that a selection rule which consults a module's path breaks the
scheme even though the channel is untouched. We verify this directly by replacing the syntactic
eligibility test with a path-based one that excludes modules by the directory they occupy, and
re-running the grid unchanged.

Under $50\%$ relocation the syntactic rule leaves the carrier set exactly unchanged, at
$160 \rightarrow 160$ on \textsc{Django}, while the path-based rule adds $31$ carriers and loses
$17$ there. The payload cost of
that difference is severe. On \textsc{Django}, mean wrong bits rise from $1.1$ to $12.6$ at $5\%$
relocation, and from $15.9$ to $34.3$ at $50\%$, past the chance floor. On a clean tree with no attack at all, the
path-based rule already fails to recover one key exactly. A rule that misclassifies a few dozen of
several hundred modules therefore costs far more than the change in carrier count alone would
explain, because the keyed reading order propagates each error forward.

\subsection{A carrier-aware adversary}
\label{sec:eval:carrieraware}

The adversary of \S\ref{sec:eval:relocation} spends its budget over every module in the repository.
Our threat model grants the adversary the scheme in full, which includes the eligibility rule, so a
keyless adversary can compute the carrier set $C$ and spend the same budget there instead.
It still cannot compute the candidate shortlists or the reading order, both of which are keyed, and
therefore cannot tell which carriers hold the payload: at a $64$-bit payload with $E=2$ only the
first $32$ carriers in the keyed order matter, and which $32$ is a function of the key. It can only
raise the density of its moves on $C$ and rely on that.

The adversary we implement targets a \emph{superset} of $C$ rather than $C$ itself: it applies the
eligibility rule up to but not including the discovery-bound exclusion, giving $211$ modules against
the encoder's $160$ on \textsc{Django}, and exactly $C$ on \textsc{Salt} and \textsc{Pylint}. On
\textsc{Pylint} it applies the \texttt{checkers} rule exactly as the encoder does, which is why its
moves never change the carrier set there. It is therefore a slightly blunter attack than the threat model
permits, and we report it as such. Each module is moved at most once.

Table~\ref{tab:carrieraware} compares the two attacks. At $5\%$ and $20\%$ they make an equal number
of realised moves, so the only difference is where those moves land; at $50\%$ they do not, and the
caption gives both counts. The budget is a fraction of the
repository's modules, and the figures are moves that succeeded, not moves attempted: a move is
abandoned when every destination already holds a file of that name. The budgets are equal at $5\%$
and $20\%$. At $50\%$ the carrier-aware adversary exhausts its carrier set and stops, at $211$ moves
against $338$ on \textsc{Django} and $412$ against $442$ on \textsc{Salt}, so it does more damage
with fewer moves than an equal-budget framing would suggest.

\begin{table}[t]
\caption{Untargeted against carrier-aware relocation at a $64$-bit payload, uncoded ($r=1$). Equal
realised moves at $5\%$ and $20\%$; unequal at $50\%$, where the carrier-aware adversary exhausts
its set. ``Pay.'' is the mean number of payload-bearing carriers relocated and ``Imp.'' the
fraction of attacked repositories that still import. Ten keys for \textsc{Django} and
\textsc{Pylint}, six for \textsc{Airflow}, five for \textsc{Salt}, three attack seeds each. One \textsc{Salt} key is excluded
because its grid did not complete within the scheduling window; no key is excluded on the ground of
its result. Budgets are equal at $5\%$ and $20\%$; at $50\%$ the carrier-aware adversary exhausts
its carrier set and stops, at $211$ moves against $338$ on \textsc{Django} and $412$ against $442$
on \textsc{Salt}. Cells are means; the standard deviation across keys and seeds runs from $0.017$ to $0.166$, is
largest on \textsc{Pylint} untargeted at $5\%$, and exceeds its own mean in one cell,
\textsc{Django} untargeted at $5\%$ ($0.098$ against $0.048$), where a single key dominates.}
\label{tab:carrieraware}
\centering
\small
\setlength{\tabcolsep}{2pt}\small
\begin{tabular}{llcccc}
\toprule
Repository & Budget & \multicolumn{2}{c}{BER} & \multicolumn{2}{c}{Pay. / Imp.} \\
\cmidrule(lr){3-4}\cmidrule(lr){5-6}
 & & untarg. & aware & untarg. & aware \\
\midrule
\multirow{3}{*}{\textsc{Django}}
 & $5\%$  & 0.048 & \textbf{0.099} & 1.6 / 0.97 & 5.7 / 0.73 \\
 & $20\%$ & 0.100 & \textbf{0.381} & 6.7 / 0.73 & 22.8 / 0.20 \\
 & $50\%$ & 0.246 & \textbf{0.517} & 16.2 / 0.27 & 32.0 / 0.00 \\
\midrule
\multirow{3}{*}{\textsc{Airflow}}
 & $5\%$  & 0.026 & 0.034 & 1.1 / 0.00 & 1.7 / 0.00 \\
 & $20\%$ & 0.101 & \textbf{0.170} & 4.4 / 0.00 & 7.7 / 0.00 \\
 & $50\%$ & 0.270 & \textbf{0.424} & 11.2 / 0.00 & 18.3 / 0.00 \\
\midrule
\multirow{3}{*}{\textsc{Salt}}
 & $5\%$  & 0.024 & 0.058 & 1.4 / 0.73 & 3.5 / 0.93 \\
 & $20\%$ & 0.076 & \textbf{0.242} & 5.8 / 0.07 & 14.2 / 0.47 \\
 & $50\%$ & 0.276 & \textbf{0.525} & 16.8 / 0.00 & 32.0 / 0.00 \\
\midrule
\multirow{3}{*}{\textsc{Pylint}}
 & $5\%$  & 0.284 & 0.052 & 3.0 / 0.90 & 6.3 / 0.77 \\
 & $20\%$ & 0.468 & 0.197 & 14.2 / 0.43 & 24.2 / 0.40 \\
 & $50\%$ & 0.459 & 0.487 & 31.3 / 0.27 & 58.9 / 0.27 \\
\bottomrule
\end{tabular}
\end{table}

\textbf{Targeting helps the adversary.} On \textsc{Django} the same
number of moves costs $0.100$ untargeted and $0.381$ carrier-aware at a $20\%$ budget, and at $50\%$
the carrier-aware attack reaches $0.517$, which is chance-level bit accuracy under this
decoder.
\textsc{Salt} reaches $0.525$ and \textsc{Airflow} $0.424$ at the same budget. The relocation
figures of
\S\ref{sec:eval:relocation} therefore describe a weaker adversary than the threat model permits,
and should be read as such.

Two measured qualifications belong with that result, and the first is not uniform. On
\textsc{Django} the attack buys its advantage by destroying the artifact: importability falls from
$0.73$ to $0.20$ at a $20\%$ budget, and at $50\%$ no attacked tree imports on \textsc{Django},
\textsc{Airflow} or \textsc{Salt}. On \textsc{Airflow} no attacked tree imports at any budget, by
either attack. \textsc{Salt} at $20\%$ runs the other way, $7$ of $15$ attacked
trees importing against $1$ of $15$ untargeted, and the reason is visible in the same rows: at
equal moves the untargeted attack relocates $176$ modules of which only $80$ are carriers, while the
carrier-aware attack relocates $176$ carriers. The carrier set is the set of modules that pass the implemented
eligibility checks, which have already excluded the discovery-bound and configuration-pinned ones,
so confining an attack to it avoids the modules those checks flag as unsafe to relocate. The checks
do not certify that every remaining relocation is safe, which is why the encoder verifies the import
graph and retries. On \textsc{Pylint} $8$ of $30$ still import at the $50\%$ budget. The joint claim
that erasing the payload also breaks the software therefore does not hold generally, and
\S\ref{sec:discussion} states where it does.

Second, the carrier set is identical before and after the attack in every carrier-aware cell on all
four repositories, so Eq.~\eqref{eq:invariance} holds under this attack; the damage is to the
symbols, not to the selection channel.

\textsc{Pylint} inverts the ordering, and the reason is visible in the same measurements. There the
untargeted attack does not merely move carriers, it changes the carrier set itself in every one of the $90$ attacked trees, where the carrier-aware attack changes it in none. The cause is one exclusion in our
implementation that is still written against the path: modules under \texttt{checkers/} are
excluded, because \textsc{Pylint} discovers its checkers by walking that directory. This is the same
hazard the syntactic rule of \S\ref{sec:carriers} handles in general, implemented for this one case
the older way. Moving a module out of \texttt{checkers/} therefore makes it eligible, and in a
$30$-move probe five modules entered the carrier set on exactly that account and none left.
$\mathcal{I}$ is not preserved, so Eq.~\eqref{eq:invariance} does not apply, and this is the cost
\S\ref{sec:eval:invariance} measures for path-based rules, appearing in the one place we had not
replaced. The keyed reading order then propagates each gained or
lost carrier through every later symbol, which is why the untargeted attack is the more damaging of
the two on this repository. This is the amplification effect of \S\ref{sec:carriers} appearing
from the attacker's side, and it is why targeting the carrier set is not uniformly the best keyless
strategy.

\subsection{A repairing adversary}
\label{sec:eval:repair}

The attack above buys its advantage at the cost of the artifact, which invites the obvious reply:
an adversary would repair the references it broke, as a maintainer performing a refactor does. We
implemented that adversary. It is the carrier-aware attack with one addition, that after the moves
every reference to a moved module is rewritten across the repository, for import statements and
attribute access, two of the three forms of \S\ref{sec:deploy}; dotted-path string literals are not
rewritten. It still holds no key.

\begin{table}[t]
\caption{The carrier-aware adversary with and without reference repair, at a $64$-bit payload,
uncoded ($r=1$). ``Imp.'' is the fraction of attacked repositories that still import. Five keys for
\textsc{Django} and \textsc{Pylint}, four for \textsc{Airflow}, three attack seeds each.}
\label{tab:repair}
\centering
\small
\setlength{\tabcolsep}{3.6pt}
\begin{tabular}{llcccc}
\toprule
 & & \multicolumn{2}{c}{BER} & \multicolumn{2}{c}{Imp.} \\
\cmidrule(lr){3-4}\cmidrule(lr){5-6}
Repository & Budget & plain & repaired & plain & repaired \\
\midrule
\multirow{3}{*}{\textsc{Django}}
 & $5\%$  & 0.138 & 0.138 & 0.67 & 0.67 \\
 & $20\%$ & 0.455 & 0.455 & 0.13 & 0.13 \\
 & $50\%$ & 0.532 & 0.532 & 0.00 & 0.00 \\
\midrule
\multirow{3}{*}{\textsc{Airflow}}
 & $5\%$  & 0.039 & 0.039 & 0.75 & 0.75 \\
 & $20\%$ & 0.190 & 0.190 & 0.17 & 0.17 \\
 & $50\%$ & 0.492 & 0.492 & 0.00 & 0.00 \\
\midrule
\multirow{3}{*}{\textsc{Pylint}}
 & $5\%$  & 0.060 & 0.060 & 0.80 & 0.80 \\
 & $20\%$ & 0.234 & 0.234 & 0.60 & 0.60 \\
 & $50\%$ & 0.537 & 0.537 & 0.13 & 0.13 \\
\bottomrule
\end{tabular}
\end{table}

``Imports'' here means that a fresh interpreter can import the package's public surface; the plain
arm does not rewrite the references it breaks. Repair
changes neither column, in all eighteen cells, and the second half of that is worth substantiating
rather than leaving as a surprise. Bit error is unchanged because repair rewrites
references and moves nothing, so the decoder reads the placements the attack left. Importability is unchanged, and in the failing trees we
inspected the residual failure is not a dangling reference but a circular import created by the
relocation itself: on \textsc{Django}, after $430$ import sites and $26$ attribute sites were
repaired, the tree still fails with

\begin{quote}\small\ttfamily
ImportError: cannot import name 'DjangoTemplates' from partially initialized module
'django.contrib.humanize.django' (most likely due to a circular import)
\end{quote}

\noindent The reference in that message is the one the repair wrote and it is correct; the cycle is
what is new, and the repair pass we implemented does not address it. Our encoder avoids it by verifying the
import graph after each round of moves and retrying (\S\ref{sec:embedding}); an adversary that adds such
a loop is the strongest attack we are aware of against this scheme and one we have not measured.

\subsection{Identity-changing transformations}
\label{sec:eval:identity}

The transformations of \S\ref{sec:eval:transforms} cost nothing because they preserve both inputs of
$\mathrm{mid}(m)$. This subsection attacks those inputs directly, to characterise the boundary of the identity
design.

\begin{table}[t]
\caption{Transformations that change the inputs of $\mathrm{mid}(m)$, at a $64$-bit payload, uncoded
($r=1$), three attack seeds per cell and five keys per repository, three for \textsc{Airflow}, on
all four repositories. ``Ids'' is the fraction of carrier identities changed and ``Imp.'' the fraction of
attacked repositories that still import. BER is the range over the four repositories.}
\label{tab:identity}
\centering
\footnotesize
\setlength{\tabcolsep}{3pt}
\begin{tabular}{llccc}
\toprule
Transformation & Extent & Ids & BER & Imp. \\
\midrule
\multirow{4}{*}{Rename carrier files}
 & $10\%$  & 0.10 & 0.36--0.49 & 0.00--0.13 \\
 & $25\%$  & 0.25 & 0.48--0.50 & 0.00 \\
 & $50\%$  & 0.50 & 0.45--0.48 & 0.00 \\
 & $100\%$ & 1.00 & 0.45--0.52 & 0.00 \\
\midrule
\multirow{4}{*}{Rename top-level definitions}
 & $10\%$  & 1.00 & 0.46--0.49 & 0.00 \\
 & $25\%$  & 1.00 & 0.43--0.48 & 0.00 \\
 & $50\%$  & 1.00 & 0.44--0.51 & 0.00 \\
 & $100\%$ & 1.00 & 0.46--0.49 & 0.00 \\
\bottomrule
\end{tabular}
\end{table}

Table~\ref{tab:identity} gives the range across the four repositories.
Both transformations drive bit accuracy to chance level, and the second does so immediately:
renaming a tenth of the top-level definitions already changes every carrier identity, because the
transformation is applied per file and renames $\max(1, \lceil 0.1\,d \rceil)$ of a file's $d$
top-level definitions, so every file that declares any gets at least one renamed, and one is enough
to change that file's identity. The fraction is therefore a floor on what the attack touches, not a
sample across the tree. File renaming is graded in the fraction renamed and reaches chance
level by a quarter; at a tenth it sits between $0.36$ and $0.49$ across the four repositories, and a
small number of attacked
trees still import at that level, where at every higher fraction none does.

The mechanism is the one \S\ref{sec:carriers} predicts. A changed identity removes a carrier from
the set and adds a different one, the keyed reading order is recomputed over a different set, and
every symbol between the positions of the removed and the added carrier is read from the wrong
carrier, and as more identities change these shifted ranges overlap and cover most of the payload. The order prefix
measured here falls from $32$ to single digits at a tenth of the files renamed, and to zero at a
quarter.

\textbf{These attacks mostly do not leave working software.} The attacked repository fails to import
in every cell of Table~\ref{tab:identity} except file renaming at a tenth, where a small number of
trees still import, because renaming a file does not rewrite the imports that name it and renaming a
definition does not rewrite its uses in other modules. An adversary that
wanted a usable artifact would have to repair all of those references, which is the work the
original maintainers would have to do for the same refactor. Identity dependence and this cost are reported together: either in isolation would misstate how
cheaply the mark can be removed.

\subsection{Runtime and memory}
\label{sec:eval:runtime}

Table~\ref{tab:runtime} reports what the scheme costs to run. Both analyses and both codec
directions complete in seconds on every repository, and peak resident memory stays under $40$\,MB,
so embedding dominates the cost: the eligibility and destination analyses together take $0.6$ to $8.7$\,s against $12$ to $70$\,s for embedding.

\begin{table}[t]
\caption{Runtime of each stage and peak resident memory, one key, $64$-bit payload, uncoded
($r=1$). Times in seconds on one core of an AMD EPYC node, memory in MB. Every run recovered the
payload exactly.}
\label{tab:runtime}
\centering
\small
\setlength{\tabcolsep}{2.6pt}
\begin{tabular}{lrrrrrrr}
\toprule
Repository & Mod. & Carr. & Eligib. & Dest. & Embed & Extr. & RSS \\
\midrule
\textsc{Pylint}  & 180 &  83 & 0.28 & 0.29 & 12.3 & 0.57 & 22.8 \\
\textsc{Airflow} & 698 & 361 & 1.19 & 0.69 & 19.0 & 1.88 & 22.7 \\
\textsc{Django}  & 864 & 160 & 0.93 & 7.08 & 31.5 & 8.00 & 18.0 \\
\textsc{Salt}    & 967 & 412 & 2.83 & 5.86 & 69.7 & 8.70 & 37.4 \\
\bottomrule
\end{tabular}
\end{table}

Embedding is the slowest stage because it verifies the import graph after each round of moves;
extraction, which performs no I/O beyond reading the tree, is $3.9$ to $21.6$ times faster.
Eligibility analysis scales with the number of modules and destination analysis with the number of
packages, which is why \textsc{Django}, with fewer carriers than \textsc{Salt} but a deeper package
tree, spends more of its time on the latter.

\subsection{Clean recovery with the raw encoder}
\label{sec:eval:wet}

Table~\ref{tab:utilisation} reports residual error on an unattacked tree for three keys of the
import-verifying encoder, which is too few to separate an unlucky key from an unreliable
utilisation. This section runs thirty keys per payload size, each carrying a payload derived from
its own key, and uses the \emph{raw} encoder, so it measures what the placement channel admits
rather than what the verifying loop can reach. It is a different configuration from
Table~\ref{tab:utilisation}, not the same experiment with more keys.

\begin{table}[t]
\caption{Exact recovery on an unattacked tree over \textbf{thirty keys} per cell, uncoded ($r=1$),
using the \textbf{raw} encoder. Every other table in this paper uses the import-verifying encoder.
Importability of the raw encoder's output is not measured here; \S\ref{sec:eval:behaviour} reports
it for the import-verifying encoder.}
\label{tab:wet}
\centering
\small
\begin{tabular}{lccc}
\toprule
Repository & Payload & Utilisation & Exact-recovery rate \\
\midrule
\multirow{4}{*}{\textsc{Django}}
 & 32\,b  & 10\% & \textbf{1.00} \\
 & 64\,b  & 20\% & \textbf{1.00} \\
 & 128\,b & 40\% & \textbf{1.00} \\
 & 256\,b & 80\% & \textbf{1.00} \\
\midrule
\textsc{Airflow} & 32--256\,b & 3--24\% & \textbf{1.00} \\
\midrule
\textsc{Salt}    & 32--256\,b & 4--31\% & \textbf{1.00} \\
\midrule
\multirow{2}{*}{\textsc{Pylint}}
 & 32\,b & 39\% & 0.90 \\
 & 64\,b & 77\% & 0.83 \\
\bottomrule
\end{tabular}
\end{table}

On \textsc{Django}, \textsc{Airflow} and \textsc{Salt} every one of the thirty keys recovered its
payload exactly at every payload size, including $256$ bits at $80\%$ utilisation on
\textsc{Django}. \textsc{Pylint} is the exception, falling to $0.83$ at $77\%$ utilisation. Its destination universe
of ten packages gives a radix of eight and, at $E=1$, four packages per symbol value, which leaves
the encoder few ways to satisfy a required symbol; we did not instrument the failing runs, so we
offer that as the likely cause rather than a measured one. Across the cells here the recovery rate
falls as utilisation rises, but repository, radix and encoder variant differ together, so we report
the association and do not attribute it to utilisation alone.

Comparing the two tables locates the residual errors of Table~\ref{tab:utilisation}. The raw
encoder does not produce them on \textsc{Django} or \textsc{Salt}, while the import-verifying
encoder does, which is consistent with their arising from placements the verifying loop refuses
rather than from the channel at that utilisation. The two tables differ in key count as well as in
encoder, so this is a consistent reading rather than an isolated cause. Both are reported, because the verifying encoder is the one that ships and the raw one shows what
this implementation writes and reads back when the verification is removed.

\subsection{Behaviour preservation}
\label{sec:eval:behaviour}

Each package's public import surface is exercised in a fresh interpreter before and after embedding,
on copies of the same tree, and passes in both columns on all four repositories. On \textsc{Django}
we also ran the project's own test suite.

\begin{table}[t]
\caption{\textsc{Django}'s own test suite before and after embedding a $64$-bit payload, uncoded
($r=1$), on copies of the same tree. ``Fail'' excludes the suite's five expected failures, and the
test count includes $1{,}626$ skipped. The payload extracts exactly from the marked tree.}
\label{tab:behaviour}
\centering
\small
\setlength{\tabcolsep}{7pt}
\begin{tabular}{lrrr}
\toprule
Tree & Tests & Fail & Err. \\
\midrule
Unmarked & 16{,}412 & 0 & 0 \\
Marked   & \textbf{16{,}412} & \textbf{1} & \textbf{0} \\
\bottomrule
\end{tabular}
\end{table}

The marked tree runs the same $16{,}412$ tests as the unmarked one, with no errors and one
additional failure, \texttt{test\_repr\_empty}, which compares the string representation of a
template engine against a literal containing a loader class's dotted path. The output does change,
and the change is the module's new dotted path. This is functional validation within the tested
configuration, not a general guarantee of behaviour preservation.

\section{Discussion and Limitations}
\label{sec:discussion}

\subsection{Identification and provenance}

\sys{} answers a narrow question. Given a repository in circulation and a key, it recovers the
payload that was embedded under that key, which identifies the distribution the copy came from. It
does not establish who owns the code. An adversary holding the same key can run an ordinary
embedding operation and obtain a repository that carries its own payload perfectly, with no residue
of the first. This is not an implementation weakness but a consequence of the property that makes
blind decoding possible: the second encoder recomputes the same carrier set in the same order, so it
overwrites rather than collides. Establishing priority therefore requires an external append-only
record of the kind used for certificate transparency~\cite{rfc6962}, and such a record settles
priority whether or not the artifact is watermarked. The payload is also not self-authenticating. A
verifier who recovers the expected bits learns that the arrangement in hand decodes to them under
that key. The payload carries no signature of its own, so this is evidence of descent only to the
extent that the key is held privately; it does not show that the key's legitimate holder placed the
bits.

\subsection{Robustness boundaries}

The robustness results in Section~\ref{sec:eval} cover different mechanisms, and a single summary
figure would hide which. We state the boundary of each.

\textbf{Transformations that leave the identity inputs intact.} Reformatting, comment stripping,
statement reordering within a file, local variable renaming, and the other transformations of
Table~\ref{tab:transforms} do not touch a carrier's file name or its top-level definition names, so
they do not touch the selection channel. All $216$ trials recovered the payload exactly
(\S\ref{sec:eval:transforms}). This is the regime the carrier is designed for, and the result is
expected rather than coincidental: such a transformation does not move a module and does not
rename one, so it leaves both inputs of $\mathrm{mid}(\cdot)$ and the placement alone. It remains a
statement about the transformations we ran, since another rewriter could touch an eligibility input
we did not exercise.

\textbf{Untargeted relocation.} An adversary that relocates files without computing the carrier set
spends most of its budget on non-carriers. On \textsc{Django}, \textsc{Airflow} and \textsc{Salt}
degradation is gradual and the payload survives substantial reorganisation
(\S\ref{sec:eval:relocation}, \S\ref{sec:eval:carrieraware}). On \textsc{Pylint} the untargeted
attack changes the carrier set through the path-based \texttt{checkers} rule, and bit error is
already $0.284$ at a $5\%$ budget.

\textbf{Carrier-aware relocation.} An adversary that recomputes a superset of the public carrier set
and spends its moves inside it is markedly more effective. At a $50\%$ budget it reaches
chance-level accuracy on \textsc{Django} ($0.517$) and \textsc{Salt} ($0.525$), against $0.246$ and
$0.276$ for the untargeted adversary at the same budget (\S\ref{sec:eval:carrieraware}). The carrier
set is public by construction, since the decoder derives it without the key, so this adversary needs
no secret. What the key protects is the mapping from carriers to symbols and the reading order, not
the identity of the carriers. \textsc{Pylint} inverts the ordering for the reason given with the
measurement, so targeting the carrier set is not uniformly the strongest keyless strategy.

\textbf{Transformations that change the identity inputs.} Renaming carrier files, or renaming their
top-level definitions, changes the input of $\mathrm{mid}(\cdot)$ and therefore the carrier's
identity. Bit error is at chance level from a quarter of the files upward, and already between
$0.36$ and $0.49$ at a tenth, over both transformations and all four repositories
(\S\ref{sec:eval:identity}). This is the scheme's sharpest boundary and it follows
from the identity we chose, which is built from names: an identity defined this way cannot survive
renaming while surviving relocation, because the two attacks change disjoint halves of the names that identity is built from. It is
the clearest reason to treat module placement as a complement to a source-text watermark rather than
a replacement for it.

\textbf{What the destructive attacks cost.} Payload destruction and loss of the artifact are
measured on the same tree in the same run. Counting individual attacked trees rather than cell
means, and counting a tree as having reached the floor when its bit error is at least $0.5$: of the
$19$ relocation trees on \textsc{Django} and the $11$ on \textsc{Salt} that reached it, none
imported, and of the $173$ identity-attack trees that reached it across all four repositories, none
imported. \textsc{Pylint} is the exception: $42$ of its relocation trees reached the floor and $16$ of those
still imported. Thirteen of those sixteen come from the untargeted attack and have a carrier set that changed under
it. Recovery and importability are therefore separable: a tree can decode to noise because the
decoder reads a different carrier set, while remaining importable because nothing its import graph
depends on was broken. The mark is removable. On \textsc{Salt}, where the adversary's set
coincides with the carrier set, every configuration we measured that removed it also delivered
software that does not load. The same holds on \textsc{Django}, but there the adversary also moved
the $51$ discovery-bound modules the encoder would never touch, so that result does not show an
attack confined to $C$ would break the software. \S\ref{sec:eval:repair} measures an adversary that
tries to avoid breaking the artifact by a different route.

\subsection{Placement feasibility and software validity}

Capacity in Table~\ref{tab:capacity} is nominal in two senses. It counts the symbols the channel
admits rather than the symbols a repository can be made to carry, and it is quoted at the maximum
symbol width $E=\log_2\varrho$, which leaves the encoder no slack. At the widths actually evaluated
here it is $320$ bits on \textsc{Django}, $1{,}083$ on \textsc{Airflow}, $824$ on \textsc{Salt} and
$83$ on \textsc{Pylint}.

Two further things reduce what is reachable. First, a destination has to be writable. Over thirty
keys the raw encoder recovers every payload exactly on \textsc{Django} and \textsc{Salt} at every
payload size we measured, including $256$ bits at $80\%$ utilisation on \textsc{Django}, while
\textsc{Pylint} falls to $0.83$ at $77\%$ utilisation (\S\ref{sec:eval:wet}). Recovery falls as
utilisation rises across those cells, but repository and radix vary with it, so we report the
association without attributing it to utilisation alone, and we did not instrument the failing runs
to establish a cause.

Second, one repository in the capacity census offers no channel at all. \textsc{idlelib} has $63$
eligible carriers and exactly one admissible destination, so the radix is one and each carrier
encodes zero bits. The precondition is a wide enough package tree, not a supply of carriers, and it
is not met by every Python project.

Behaviour preservation is validated at two rungs (\S\ref{sec:eval:behaviour}). The package's own
import surface passes before and after embedding on all four repositories, and \textsc{Django}'s own
suite runs its baseline $16{,}412$ tests with no errors and one failure, an assertion on a module's
dotted path. That suite result is from one repository and bounds what it supports: the suite completed with no errors and one
additional failure, tied to a module's new dotted path, in the configuration we tested. It is not a
general guarantee of behaviour preservation.

\subsection{Scope and evaluation limits}

The scheme applies to languages in which a module's location is a free choice. We evaluate Python
only. The design appears to transfer to JavaScript and TypeScript, where placement is likewise
decoupled from semantics, but we have not measured them. Java and Go bind directory to package
declaration, so relocating a file there requires editing the file, and the channel loses the
property that makes it interesting.

Three further limits bound what the numbers support. The census covers ten repositories, and the
experiments cover a subset that varies by experiment: the embedding-cost and text-transform grids
run on \textsc{Django}, \textsc{Salt} and \textsc{Pylint}, the payload-size and coding grids on
\textsc{Django} alone, and the runtime measurement, the carrier-aware and identity attacks and the
clean-recovery grid on all four, and the repair experiment on \textsc{Django}, \textsc{Airflow} and
\textsc{Pylint}. The capacity census covers ten. Each table names its own repositories. We report every repository separately and never average across them, so each figure is
a statement about one project rather than about Python. The adversary
models are fixed strategies, including the carrier-aware one; we do not evaluate an adaptive
adversary that learns from repeated extraction attempts. And we do not evaluate whether the
placements the encoder chooses look natural to a human reviewer, which is the relevant notion of
imperceptibility for this carrier and which a measurement of import-graph validity does not
establish.

Three further boundaries follow from what our adversaries do. They only relocate existing modules.
We do not evaluate additions or deletions, and a new eligible module, an empty package or an unused
top-level definition each changes the carrier set, the destination universe or a carrier's identity,
which the keyed reading order and the shortlists would then propagate. Such edits preserve behaviour
and are ordinary development, so this is a boundary we have not measured. We also assume the
adversary holds no unmarked copy: a public upstream, two differently marked copies, or the embedding
commit in version history would each reveal the moved modules directly. And relocation changes a
module's public dotted path, which code outside the repository that imports it would observe.

\section{Conclusion}

We have shown that module placement is a usable watermark carrier for source code. It holds hundreds
to thousands of bits on production repositories, is recovered blindly from the artifact, the key
and the agreed decoding parameters, cost no payload bit across the $216$ measured trials of the code transformations reported to
damage content-carrier schemes, and degrades gradually rather than abruptly under rearrangement of
the tree on the repositories whose carrier set the rearrangement leaves intact. Recovery from an unmodified repository is exact when the selected payload positions are
writable; positions the encoder cannot write are unknown to the decoder and leave residual errors as
utilisation rises. The design
constraint that makes it work is a single property: the carrier set and reading order that the
decoder recomputes must be invariant under the operation the scheme must survive. We have shown both
that this property determines the design and what its violation costs.

The boundaries are as sharp as the result. An adversary that recomputes the public carrier set, or
in our \textsc{Django} runs a superset of it, and concentrates its moves there reaches chance-level
accuracy at a $50\%$ budget on \textsc{Django} and \textsc{Salt}. On \textsc{Salt}, where that adversary's set coincides with the carrier set, every
attacked tree that reached the floor also failed to import. Renaming carrier files or their
top-level definitions changes the identity input and drives bit
accuracy to chance level from a quarter of the files upward, which follows from deriving identity
from content and is the reason we present module placement as a complement to a source-text
watermark rather than a replacement for it. The two carriers have complementary attack surfaces
rather than disjoint ones: reformatting and local renaming leave placement untouched, relocation
leaves the code text untouched, and a refactor that renames files while reorganising them reaches
both. Neither carrier alone covers the range.

\bibliographystyle{IEEEtran}
\bibliography{refs}

\end{document}